# Role of rf electric and magnetic fields in heating of micro-protrusions in accelerating structures.

Gregory S. Nusinovich and Thomas M. Antonsen, Jr.

*Abstract*

It is known that high-gradient operation in metallic accelerating structures causes significant deterioration of structure surfaces that, in turn, greatly increases the probability of microwave breakdown. At the same time, the physical reason for this deterioration so far is not well understood. In the present paper, the role of two effects is analyzed, viz. (a) the microwave heating caused by penetration of the rf magnetic field into microprotrusion of a radius on the order of the skin depth and (b) the Joule heating caused by the field emitted current, i.e. the effect of the rf electric field magnified by a sharp protrusion. Corresponding expressions for the power densities of both effects are derived and the criterion for evaluating the dominance of one of these two is formulated. This criterion is analyzed and illustrated by the discussion of an example with parameters typical for recent experiments at the Stanford Linear Accelerator Center (SLAC) National Accelerator Laboratory.

In the development of linear accelerators of the next generation, one of the main problems is providing reliable operation with low breakdown rate using the highest gradients possible. Theoretical and experimental studies aiming at improving our understanding of possible causes of breakdown at high gradients has been an area of active research for a long time. More than a decade ago it was widely accepted that the key factor limiting high-gradient operation is field emission from small protrusions, since this field emission causes the appearance of the dark current which, for a number reasons, can be considered a precursor of the breakdown [1]. This field emission described by the Fowler-Nordheim law is caused by the rf electric field which can be greatly magnified by a sharp protrusion. Later, significant attention was also paid to how the rf magnetic field causes deterioration of accelerating structures [2]. Experimental studies of both effects are in progress (see, e.g., recent papers [3, 4]). These experiments have stimulated corresponding theoretical studies (see, e.g. [5, 6]).

As a rule, the authors focus their attention primarily either on the effect of the rf electric field or on the effect of the rf magnetic field. (Ref. 6 is an exclusion of this rule.) At the same time, it seems reasonable to believe that the role of these fields depends on experimental conditions, viz. under one set of conditions the rf electric field plays a dominant role, while under another set of conditions the most important is the rf magnetic field. Below, we analyze these contributions for the case of small cylindrical microprotrusion which may exist on surfaces of accelerating structures after a certain period of high-gradient operation.

Consider a thin cylindrical microprotrusion on a structure surface. Then, the power (per unit length) of microwave losses caused by the rf magnetic field penetrating into protrusion can be given as [7]:

$$p_{skin} = \omega \alpha_m'' S_\perp \frac{\mu_0 |H|^2}{2}. \tag{1}$$

In (1), $\omega$ is the wave frequency, $\alpha_m''$ is the imaginary part of the magnetic polarisability of a cylinder, $S_\perp$ is the cross-sectional area of the protrusion, $\mu_0$ is the permeability of free space (below we consider non-magnetic materials) and $H$ is the amplitude of the rf magnetic field. Note that in (1) we took into account only the microwave losses caused by the penetration of the rf magnetic field because estimates show [8] that under all reasonable conditions these losses are much higher than contribution from the imaginary part of the electric polarisability. Introducing the skin depth $\delta$ defined by a known formula [9] $\delta^2 = 2/\omega\mu_0\sigma$ (where $\sigma$ is the conductivity of the metal) allows us to rewrite Eq. (1) as

$$p_{skin} = \frac{1}{\sigma}\frac{S_\perp}{\delta^2}\alpha_m''|H|^2. \quad (2)$$

Expressions for the magnetic polarisability of a cylindrical wire of a circular cross-section can be found elsewhere [7- 9]. For other configurations of micro-protrusions they can be derived by using the method described elsewhere [7].

Let us now evaluate the power of Joule heating of this protrusion by an electron current. As a rule, it is assumed that, when this current is emitted from the protrusion apex, its density obeys the Fowler-Nordheim equation. In general, there are some reasons [10] to take into account also thermal contributions defined by the Richardson-Dushman equation to this current, but below we will discuss this current just bearing in mind the field emission defined by the Fowler-Nordheim law.

In general, the power of Joule heating can be defined by the standard equation $P = R\langle I^2 \rangle$ where angular brackets denote averaging over the rf period as well as possible other distributions over the angular position of an emitting site on apex, initial velocities etc (those distributions we will neglect below). Taking into account that the resistance per unit length is equal to $R = \rho/S_\perp$ where $\rho = 1/\sigma$, we can define the power (per unit length) of Joule heating caused by this electron current as

$$p_J = \frac{1}{\sigma}\frac{1}{S_\perp}\langle I^2 \rangle. \quad (3)$$

From comparison of Eq. (3) with Eq. (2) it follows that the role of the rf magnetic field is more important in protrusion heating than the role of the rf electric field causing emission of the dark current when the following condition holds:

$$\left(\frac{S_\perp}{\delta^2}\right)^2 \alpha_m'' > \frac{\langle I^2 \rangle}{\delta^2|H|^2}. \quad (4)$$

For protrusions of a circular cross-section with radius $a$, Eq. (4) reduces to

$$\Phi\left(\frac{a}{\delta}\right) \equiv \pi^2 \left(\frac{a}{\delta}\right)^4 \alpha_m'' > \frac{\langle I^2 \rangle}{\delta^2 |H|^2} \tag{5}$$

The right-hand sides of Eq. (4) and Eq. (5) contains the electron current squared averaged over the rf period $\langle I^2 \rangle$. The distribution of $I^2$ over the rf period has been analyzed elsewhere [6, 11]. It was shown [11] that for typical values of the work function 4-5 eV and maximum surface gradients not exceeding 10 GV/m, the term $\langle I^2 \rangle$ is less than 0.1 of its peak value $I^2$. The latter can be treated as its value in the DC electric field $(I^{(0)})^2$. Below we will denote the ratio $\langle I^2 \rangle / I^2$ by $\Psi$ and present $\langle I^2 \rangle$ as $\Psi (I^{(0)})^2$.

It should be noted that the DC component of the current $I^{(0)}$ is equal to the current emitted from the protrusion apex. The electron current density on the apex obeys the Fowler-Nordheim law that takes into account magnification of the rf electric field by a sharp end of the protrusion. Then, the total emitted current is the product of this density and the apex area. Typically, the area of this apex is much smaller than the cross sectional area of a central part of the protrusion. So the current density in this central part of protrusion is much smaller than on apex, but the total current remains the same.

Let us now discuss the imaginary part of the magnetic polarisability. The magnetic polarisability of a cylinder with a circular cross-section with respect to the magnetic field perpendicular to the cylinder axis can be defined as [7]

$$\alpha_m = \frac{1}{2\pi}\left[\frac{2}{ka}\frac{J_1(ka)}{J_0(ka)} - 1\right] \tag{6}$$

where $k = (1+i)/\delta$ is the complex wave number. Corresponding dependence of the imaginary part of this polarisability on the $a/\delta$-ratio is given in Ref. 8 where pulse heating by this rf field was studied with the account for the temperature dependence of conductivity and, hence, the skin depth.

In the case when the protrusion radius $a$ is much smaller than the skin depth $\delta$, the imaginary part of this polarisability can be approximated as [7]

$$\alpha_m'' = \frac{1}{8\pi}\left(\frac{a}{\delta}\right)^2. \tag{7}$$

So the Poynting flux per unit length in such a case is equal to

$$p_{skin} = \frac{1}{8\sigma}\left(\frac{a}{\delta}\right)^4 |H|^2 \qquad (8)$$

and the condition of dominance of the effect of rf magnetic field given above by (5) reduces to

$$\Phi_1\left(\frac{a}{\delta}\right) = \lim_{a/\delta \to 0} \Phi \equiv \frac{\pi}{8}\left(\frac{a}{\delta}\right)^6 > \Psi \frac{[I^{(0)}]^2}{\delta^2 |H|^2} \qquad (9)$$

This equation contains the sixth power of the radius-to-skin depth ratio. Partly, this originates from the fact that in the case of small radius of protrusion the losses caused by penetration of the rf magnetic field into it decrease with the conductor radius proportionally to the fourth power of $a$ where $a^2$ comes from the conductor area (the rf field does not "see" a very thin conductor). Another $(a/\delta)^2$ term comes from the dependence of the imaginary part of the magnetic polarisability on the radius. At the same time, as the protrusion radius gets smaller, its resistance gets larger, thus, joule heating of the conductor by the field emitted current increases.

In the opposite limiting case, when the protrusion radius is much larger than the skin depth, the imaginary part of the magnetic polarisability can be approximated [7] as $\alpha_m'' = (1/2\pi)(\delta/a)$. Thus, instead Eq. (9) the following condition holds:

$$\Phi_2\left(\frac{a}{\delta}\right) = \lim_{a/\delta \to \infty} \Phi \equiv \frac{\pi}{2}\left(\frac{a}{\delta}\right)^3 > \Psi \frac{[I^{(0)}]^2}{\delta^2 |H|^2} \qquad (10)$$

Dependencies of all three functions, the general one $\Phi(a/\delta)$ for arbitrary $a/\delta$-ratios in Eq. (5), the function $\Phi_1(a/\delta)$ for small $a/\delta$-ratios in Eq. (9) and the function $\Phi_2(a/\delta)$ for large $a/\delta$-ratios in Eq. (10) are shown in Fig. 1. As one can see, the small $a/\delta$-ratio approximation works well up to $a/\delta$ values about one, while the large $a/\delta$-ratio approximation gives reasonable agreement starting from $a/\delta$ close to 2.5.

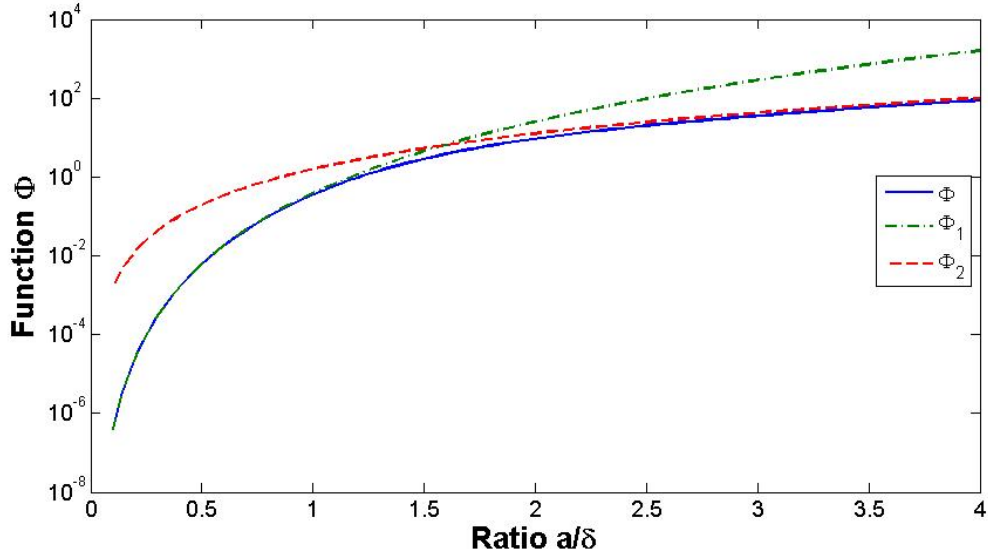

Figure 1. The function $\Phi$ and its approximations as functions of the $a/\delta$-ratio

Let us illustrate these simple formulas with some examples. Consider the operation at the 11.424 GHz frequency which for copper yields the skin depth of 0.63 micron. Let us limit our consideration by the case of thin protrusions (small $a/\delta$-ratio). Then, the range of protrusion radii we consider will be from 0.1 to 1 micron. Assume that the ratio of the averaged to peak values of the squared current $\Psi$ is close to 0.1. Now, we are left in (9) with only two parameters: the total value of the Fowler-Nordheim DC current and the intensity of the rf magnetic field. As follows from the analysis of the Fowler-Nordheim current in microprotrusions given in Ref. 10 and supported by some experimental data from SLAC, the total dark current field emitted from typical microprotrusions can be as high as 0.1 mA. Corresponding current densities from small apexes of these protrusions are on the order of a fraction of $A/\mu m^2$. Regarding the possible level of intensity of the rf magnetic field let us note that recent experiments at SLAC [12] were carried out at surface magnetic field values on the order of 0.5-0.6 MA/m. If we assume the total dark current is equal to 0.1 mA and the rf magnetic field is equal to 0.5 MA/m, we readily get from Eq. (9) that penetration of the rf magnetic field is the dominant factor for heating of microprotrusion with the radius exceeding $0.05\delta$, i.e. larger than 30 nm.

Note that to have intense heating not only should the absorbed power be large enough, but also the pulse duration should be short enough to avoid heat sink from protrusion to the body of a metallic structure. The latter means that the height of a microprotrusion $h$ should be larger than the heat propagation distance $l_h = \sqrt{D\tau}$ (here $D$ is the diffusion coefficient and $\tau$ is the microwave pulse duration). For copper the diffusion coefficient is close to $0.12\ \mu m^2/ns$. For example, in the case of 100 ns pulses, this condition $h > l_h$ is valid for microprotrusions with a height exceeding 4 microns. Therefore we should expect significant heating caused by the skin effect in the case of micro-protrusions with a height-to-radius ratio on the order of 10. Note that just such thin and long micro-protrusions were found responsible [13] for the dark current in experiments with the DC field.

Formation of such microprotrusions on the originally smooth, well polished surface of an accelerating structure is, in turn, an area of active research. In this regard, first, let us note that often the footnotes of breakdown events are found in the parts of accelerating structures where both rf electric and rf magnetic fields are present (see, e.g., Ref. 14). Then, Ref. 5 should be mentioned where surface heating by the rf magnetic field up to its melting and subsequent formation of Taylor cones is analyzed. Clearly, when the shape of such conical protrusion sharpens, this makes the radius of protrusion (at least of its part close to the apex) smaller. So, it may happen that in the course of this evolution of the protrusion shape the distribution of roles changes: while in the initial stage the key role belongs to the rf magnetic field, then, in the case of increasing height-to-base ratios, the rf electric field becomes more important. This issue of dynamic evolution of the protrusion shape, however, goes beyond the scope of the present paper.

This work was supported by the Division of High Energy Physics of the U.S. Department of Energy. The authors are indebted to S. Tantawi for insightful discussion and D. Kashyn for help in preparation of the figure.